# Modelling of tunnelling currents in Hf-based gate stacks as a function of temperature and extraction of material parameters

Andrea Campera, Giuseppe Iannaccone, *Member, IEEE,* Felice Crupi, *Member, IEEE*

A. Campera and G. Iannaccone are with the University of Pisa, Dipartimento di Ingegneria dell'Informazione: Elettronica, Informatica, Telecomunicazioni, Via Caruso 16, 56122, Pisa, Italy (phone: +39 0502217677; fax: +39 0502217522; e-mail: andrea.campera@iet.unipi.it, g.iannaccone@iet.unipi.it).

F. Crupi is with Dipartimento di Elettronica, Informatica e Sistemistica, Università della Calabria, I-87036 Arcavacata di Rende (CS), Italy (email: crupi@unical.it).

*Abstract*—In this paper we show that through electrical characterization and detailed quantum simulations of the capacitance-voltage and current-voltage characteristics it is possible to extract a series of material parameters of alternative gate dielectrics. We have focused on $HfO_2$ and $HfSi_XO_YN_Z$ gate stacks and have extracted information on the nature of localized states in the dielectric responsible for a trap-assisted tunneling current component and for the temperature behavior of the I-V characteristics. Simulations are based on a 1D Poisson-Schrödinger solver capable to provide the pure tunneling current and Trap Assisted Tunneling component. Energy and capture cross section of traps responsible for TAT current have been extracted.

## I. INTRODUCTION

THE present research and development effort focused on replacing $SiO_2$ and SiON gate stacks with high dielectric permittivity (high-*k*) gate stacks is motivated by the need of increasing gate capacitance per unit area as device size shrinks while at the same time maintaining the leakage tunneling current under control. Among high-*k* dielectrics, $HfO_2$ and its silicates - in particular hafnium silicate nitride ($HfSi_XO_YN_Z$) - are considered promising alternative gate dielectrics for CMOS technology at the 45 nm node and beyond [1,2,3]. In fact, as it is well known, high-*k* dielectrics exhibit strongly suppressed gate tunneling current with respect to silicon oxide for the same gate capacitance: of a factor $10^5$ for $HfO_2$, and a factor $10^3$ for an $HfSi_XO_YN_Z$ stack [1]. Both materials have a rather large band gap, about 6 eV for $HfO_2$ and 7 eV for $HfSi_XO_YN_Z$ [4,5]; band gap is lower with respect to conventional $SiO_2$ but the increased thickness, in any case, causes an overall leakage current suppression.

Although significant reduction of gate leakage has been achieved, the nature of the stray currents and the frequently observed flat band voltage shift and voltage bias instability [1,6,7] need further investigation. . The effect of polysilicon depletion, which will require the use of metal gates, and the reduced effective channel mobility are other important issues, which will become more critical with the shrinking of the physical dimensions.

The main aim of this work is the understanding of the tunneling mechanisms through such dielectrics. In order to do this we have investigated the current-voltage characteristics as a function of temperature. We show that the temperature dependence of I- V characteristics in $HfO_2$ can be explained in terms of a microscopic model of Trap Assisted Tunneling (TAT) based on a temperature dependent cross section. Comparison of experiments with the results of our model allows us to exclude that relevant traps for transport are localized at the $HfO_2$ interfaces (confirming some experiments [8]).

The paper is organized as follows: in Section II we describe the structures under investigation. In Section III we briefly discuss the Poisson-Schrödinger solver and how the tunneling current components are obtained. In Section IV and V we compare experimental and theoretical C-V and J-V characteristics, respectively, and extract the relevant physical parameters of the materials. In Section VI we discuss in some more detail the transport model we have used to explain the temperature dependent transport in $HfO_2$ and $HfSi_XO_YN_Z$ and the comparison with experiments, and finally we will draw our conclusions.

## II. EXPERIMENTS

We have experimentally investigated $HfO_2$ and $HfSi_XO_YN_Z$ gate stacks; the physical parameters of Hf-based dielectrics depend on the fabrication process and on the specific resulting composition. Indeed, for example, the dielectric constant of $HfSi_XO_YN_Z$ depends on the nitrogen and hafnium molar fraction in the high-*k* layer [2,9]. Therefore is important to extract with sufficient precision the physical parameters of the high-*k* layer, as we shall do by comparing experiments with simulations. We have considered three different gate stacks, consisting of an interfacial layer (we assume it to be $SiO_2$ for the $HfO_2$ gate stack and SiON for the $HfSi_XO_YN_Z$ gate stack) of thickness $t_{int}$ in contact with the silicon substrate, and of an high-*k* layer of thickness $t_{high-k}$ covered by a n+ poly-silicon gate. The gate dielectrics have been deposited by MOCVD after an IMEC-clean process. In all three cases the substrate is p-doped with $N_A=5 \cdot 10^{23}$ m$^{-3}$. Details on the layer structures are shown in Table I.

Capacitance-voltage characteristics have been measured for capacitors of area 70 μm x 70 μm and current-voltage characteristics have been measured for n-MOSFETs with width 10 μm and gate length 5 μm.

## III. 1-D POISSON-SCHRÖDINGER SOLVER

In thin Equivalent Oxide Thickness (EOT) MOS structures, such as those we are considering, polysilicon depletion and finite density of states in the bulk must be taken into account. To simulate these structures we have used a one-dimensional self-consistent Poisson-Schrödinger solver that takes into account quantum confinement at the emitting region (being it the polysilicon gate or the substrate, depending on bias), mass anisotropy in silicon conduction band, light and heavy holes, wave function penetration in the oxide. For simplicity, bands are parabolic. The band profile is computed with the quasi-equilibrium approximation, i.e., assuming that the tunneling current is so low that the dielectric separates two regions in local equilibrium with two different Fermi energies.

From the band profile (e.g. the one shown in Fig. 1), we can compute the tunneling current per unit area, that is given by [10]:

$$J = \frac{2qkT}{\pi\hbar^2} m_t \sum_i \nu_{ril} T(E_{il}) \ln\left\{\frac{1+\exp\left[(E_{Fl}-E_{il})/kT\right]}{1+\exp\left[(E_{Fr}-E_{il})/kT\right]}\right\} + \\ + \frac{4qkT}{\pi\hbar^2}\sqrt{m_t m_l} \sum_i \nu_{rit} T(E_{it}) \ln\left\{\frac{1+\exp\left[(E_{Fl}-E_{it})/kT\right]}{1+\exp\left[(E_{Fr}-E_{it})/kT\right]}\right\} \qquad (1)$$

where $q$ is electron charge, $\hbar$ the reduced Planck constant, $k_B$ the Boltzmann constant, $T$ the absolute temperature, $m_t$ ($m_l$) the transversal (longitudinal) electron effective mass, $T(E_{il})$ ($T(E_{it})$) the transmission coefficients for the longitudinal (transversal) effective mass, $\upsilon_{ril}$ ($\upsilon_{rit}$) the attempt frequencies for the longitudinal (transversal) effective mass at the emitter computed by taking into account the time spent by the electrons in the classically forbidden regions for each eigenfunction [11], $E_{il}$ ($E_{it}$) the longitudinal (transversal) eigenvalues and $E_{Fl}$ ($E_{Fr}$) the Fermi level of the left (right) electrode. The transmission coefficients are obtained by solving the Schrödinger equation for the barrier with open boundary conditions.

## IV. CAPACITANCE-VOLTAGE SIMULATIONS

It is a well known fact that, with respect to the ideal $SiO_2$ case, using an Hf-based gate stack a threshold voltage shift $\Delta V_T$ is observed for both n-MOS and p-MOS transistors [6,7,12]. This threshold voltage shift is positive for n+ poly-Si and negative for p+ poly-Si gate and is caused by defects at the poly-Si/high-$k$ interface located in the upper part of the band gap. $\Delta V_T$ is larger for p-MOS transistors with respect to n-MOS transistors. Hobbs and coworkers have explained this effect in terms of Fermi level pinning [6,7].

In Fig. 2 we compare the experimental C-V characteristics of structure *a* with those obtained from quantum simulations. Fermi-level pinning can be simply and effectively taken into account by assuming an effective electron affinity of the gate electrode, in general different from that of the gate material. The effective variation of the electron affinity can be obtained by shifting the theoretical C-V curve along the voltage axis until it overlaps with the experimental one. For structure *a* (hence for a generic $HfO_2$ stack, since Fermi level pinning is an interfacial effect and does not depend on the thickness of the $HfO_2$ layer) we have found such quantity to be 0.35 V, in good agreement with values reported in literature ( e.g. about 0.3 V in [6,7]). Interfacial charge and charge within the high-k layer can, in the same manner, contribute to the $V_T$ shift. However in Ref. [7] a detailed investigation of the charge in the various gate stack layers is presented and the authors have observed that almost the entire shift is caused by FLP. The Fermi level pinning reported in Ref. [7] depends on the hafnium molar fraction and is between 0.2 V and 0.35 V. Given the lack of knowledge fixed charges in the stack we assumed that all the $V_T$ shift was entirely caused by the FLP at the poly-high-k interface. The value that we extracted is in good agreement with findings in Ref. [7] hence we assumed that the total $V_T$ shift was caused by the FLP. It is important to note that in Fig. 2 and Fig. 3 we compare experimental C-V characteristics measured at high frequency with theoretical low-frequency C-V characteristics: hence the discrepancy at positive voltages. From the comparison between experimental and simulated C-V characteristics we can also extract the dielectric constant. We find a value of 25 for the relative dielectric constant of the hafnium oxide in good agreement with values reported in literature (for example [4]). In addition, from Fig. 2 we can observe a smoother experimental C-V characteristic with respect to simulations, which means that the effect of interfacial traps is significant. We want to remark that we are assuming that the layer thicknesses are known with sufficient precision.

We can repeat the same procedure for the $HfSi_XO_YN_Z$ gate stack (structures *b* and *c*) in order to extract the dielectric constant and the Fermi level pinning. In Figure 3 we report the C-V curves with and without FLP for structure *b*. Fermi level pinning extracted from the comparison is of 0.13 eV, that agrees with values reported in literature [5,6,8]; the dielectric constant of the $HfSi_XO_YN_Z$ results to be 11 (very close to [8]). The very similar slope of the experimental and theoretical C-V profiles in Fig. 3 also shows that $HfSi_XO_YN_Z$ has less interfacial traps.

## V. Current Density-Voltage Simulations

From the comparison between theoretical and experimental current density-voltage characteristics, we are able to extract the electronic affinity and the electron tunneling effective mass for $HfO_2$ and $HfSi_XO_YN_Z$, and to evaluate the presence of transport mechanisms additional to pure elastic tunneling. The electron affinity and the effective mass have a slightly different effect on the J-V characteristics. Exploring the parameter space we could only find a set of values that allow us to fit both the J-V and the C-V characteristics for positive and negative gate voltages. In Fig. 4 we plot experimental and theoretical (pure tunneling) currents for $HfO_2$ and $HfSi_XO_YN_Z$ stacks, for negative and positive gate bias. The extracted electron affinity and electron effective mass, are, respectively, 1.75 eV and 0.08 $m_0$ for $HfO_2$, 1.97 eV and 0.24 $m_0$ for $HfSi_XO_YN_Z$ that have to be compared with values reported in literature [4,5,9,13,14].

Hole tunneling is not significant: we have used a tunneling mass for holes of $0.5m_0$ (see for example [15]), although such value is somewhat arbitrary, since it cannot be verified in a straightforward way. For the electron tunneling effective mass in $SiO_2$ we have used $0.5m_0$. The situation is more complicate for the SiON layer for which the electron effective mass, and all the other physical parameters, depend on the molar fraction of nitrogen embedded in the oxide: we have used a value of $0.45m_0$, an electron affinity of 1.27 eV and a dielectric permittivity of 5. Such values allow us to fit the experiments, and are in good agreement with values reported in Ref. [16]. Extracted values for material parameters are summarized in Table II.

We want to stress the fact that from Fig. 4 it is clear that for the HfSi$_X$O$_Y$N$_Z$ stack (structure *c*), and for the HfO$_2$ stack (structure *a*), the pure tunneling component is sufficient to fit the whole J-V characteristics except for low negative gate bias between -1 and 0 V for structure *c*, and between -2 and 0 V for structure *a*. The higher measured current density can be ascribed to two effects: an additional transport mechanism assisted by interface traps in the silicon gap [17], and an higher current density associated with the lateral source and drain n+ extensions, not considered in the 1D simulations. Consequently, an higher overall gate current density is measured in shorter channel length n-MOSFETs at low negative gate voltages, as shown in Fig. 5.

## VI. TEMPERATURE DEPENDENCE OF CURRENT-VOLTAGE CHARACTERISTICS

HfO$_2$ and HfSi$_X$O$_Y$N$_Z$ show a different behavior as a function of temperature as we can see in Figs. 6 and 7, respectively. Current in HfSi$_X$O$_Y$N$_Z$ stack is substantially independent of temperature, whereas in HfO$_2$ a larger temperature dependence is present for positive gate voltages, lower for negative gate voltages. We explain this behavior by introducing a temperature dependent cross section in a microscopic model of Trap Assisted Tunneling. Here we briefly report the basic physics of the model, addressing the interested readers to the original work [19], and describe in some more detail the features that we have introduced to take into account the temperature dependence.

Let us consider the band profile sketched in Fig. 8, representing structure *a* with a positive gate applied voltage of 0.5 V, an electron state $|\beta\rangle$ in one band of one electrode and a Khon Sham state $|\alpha\rangle$ representing a trap state in the dielectric stack. The probability per unit time that an electron tunnel from one band of one reservoir to the trap is given by the Fermi golden rule and is:

$$v_{\beta \to \alpha} = \frac{2\pi}{\hbar} |M(\beta,\alpha)|^2 h_\Gamma (E_\beta - E_\alpha) \qquad (2)$$

where

$$h_\Gamma (E_\beta - E_\alpha) = \frac{1}{\pi} \cdot \frac{\Gamma}{(E_\beta - E_\alpha)^2 + \Gamma^2} \qquad (3)$$

and E$_\beta$, E$_\alpha$ are the energy of the state $|\beta\rangle$ and $|\alpha\rangle$ respectively. In (2), we take into account for inelastic transitions by using a a Lorentzian function of finite half width $\Gamma$. We can rewrite (2) as

$$v_{\beta \to \alpha} = \sigma_{\beta,\alpha} J(\beta,x) = \sigma_{\beta,\alpha} \cdot T(E_l) \cdot f(E_l) \qquad (4)$$

if we define the energy dependent capture cross section as

$$\sigma_{\beta,\alpha} = \sigma_0 \cdot h_\Gamma (E_\beta - E_\alpha) \qquad (5)$$

$J(\beta,x)$ is the probability current density impinging on the plane positioned at the same depth $x$ of the trap, while $\sigma_0$, expressed as m²·J, is a compact parameter that contains all information on traps. We want to stress the fact that our definition of capture cross section is slightly different from the conventional one, and allows us to introduce a dependence of the transition rate upon the energy difference between initial and final states.

Once we know the band profile we can compute the capture and emission rates (which we will call in the rest of this paper generation and recombination rates, respectively), i.e. the probability per unit time that an electron tunnels from one band of one electrode to the trap and vice versa, and then the trap assisted tunneling current density.

$$g = 2\int_\beta v_{\beta \to \alpha} \rho_\beta f_\beta d\beta \tag{6}$$

$$r = \int_\beta v_{\beta \to \alpha} \rho_\beta (1-f_\beta) d\beta \tag{7}$$

We can also write the TAT current density as [19]:

$$J = q \cdot \frac{g_1 r_2 - g_2 r_1}{g_1 + g_2 + r_1 + r_2} \tag{8}$$

where subscript 1 of the generation and recombination rates refers to states $\beta$ in the substrate and subscript 2 to states $\beta$ in the gate electrode.

The TAT current density temperature dependence can be included in an appropriate model for $\sigma_{\beta,\alpha}$ that includes its temperature dependence.

Traps in hafnium oxide have been recently investigated by Gavartin and coworkers [20] with *ab-initio* calculations. From simulations we have observed that traps must be within the energy range 1÷2 eV below the hafnium oxide conduction band in order to allow us to reproduce the shape of the J-V characteristics at various temperatures. In Ref. [20] only one defect is in the energy range above, and is specifically located 1.6 eV below the hafnium oxide conduction band ($O^0/O^-$ defects), hence we assume them to be the dominant traps for the TAT current. In Fig. 9 we report the band profile for structure *a* and the energetic position of the considered traps.

In order to simulate the TAT current we need to know the half width of the Lorentzian function $\Gamma$, and the capture cross section. We can extract $\Gamma$ from the slope of the J-V curve at 475 K (see Fig. 10). Indeed we can reasonably assume, and this is confirmed by simulations, that at 475 K the TAT current represents the entire current density because the pure tunneling current depends only slightly on temperature.

It is important to note that $\Gamma$ strongly affects the slope of the J-V characteristics on the semilog scale whereas $\sigma_0$ appears as a multiplicative factor in $J$ (Eq. 8) so that a variation of $\sigma_0$ only shifts vertically the J-V characteristics on the semilog scale. On the other hand, from the experimental results shown in Fig. 6, we can see that the J-V characteristics have a slope almost independent of temperature, and therefore we can assume that in our case Γ is practically constant with temperature in the operating regimes we consider.

We now have to consider the temperature dependence of $\sigma_0$, which is typically given by an Arrhenius-like behavior [21]. We shall assume:

$$\sigma_0 = \sigma_\infty \exp(-E_\sigma/k_B T) \tag{9}$$

where $\sigma_\infty$ is the capture cross section for $T \rightarrow \infty$, $E_\sigma$ is the activation energy of the capture process.

For each temperature we determine the value of $\sigma_0$ which provides the best fit between the theoretical and experimental J-V characteristics. The extracted $\sigma_0$ is plotted in Fig. 11 as a function of temperature: as can be seen, it perfectly fits an Arrhenius function with an activation energy $E_\sigma$=0.542 eV and $\sigma_\infty$ = 0.555 m$^2$J. We should note that the activation energy of the capture cross section is rather high with respect to that typically found for SiO$_2$, which is lower than 0.3 eV [22], but to our knowledge no other data are available for hafnium oxide, and we can not exclude *a priori* this value. Indeed some materials, such as *a* Si:H, exhibit an activation energy even higher than that we have obtained (0.69 eV [23]).

With the extracted values of Γ and σ(*T*) the TAT current can be well reproduced for positive gate voltage, as sketched in Fig. 12. For negative gate applied voltage TAT current is less relevant because the traps have a much higher energy with respect to the emitter Fermi level, as we can see in Fig. 13. In addition, we must remark that an additional transport mechanism associated with the lateral source and drain n+ extensions, is present for relatively "low" negative gate voltage (-1<V$_G$<0) and this is not considered in our 1D simulations. (see Fig. 5). Then we have to focus on higher negative gate voltages (-1.6<V$_G$<-1).
The agreement for negative gate applied voltages between theoretical and experimental curves is reasonable, but not as good as for positive voltages mainly because the low absolute values for the current density leads to an increased relative error, as shown in Fig. 14. For the hafnium silicate nitride experiments show a very small temperature dependence, which can be explained in terms of pure tunneling current. We cannot exclude that traps are present in the hafnium silicate nitride bulk, but we point out that are not responsible for a significant trap-assisted tunneling component. In Fig. 15 experiments and simulations at 298 and 400 K are reported for structure *c*: they agree pretty well without any additional fitting parameter.

The temperature dependent TAT model allows us to reproduce experiments if we assume that traps in hafnium oxide are uniformly distributed in the layer. We have checked that if we assume that traps are placed exclusively at one of the hafnium oxide interfaces, we are not able obtain a good agreement with measurements. Indeed, in this latter case, the slope of the theoretical curves would depend on temperatures, while that of the experimental curves would not. Our assumption that traps are uniformly distributed in the hafnium oxide layer is also confirmed by some recent experiments [8].

## VII. Conclusion

We have shown that the temperature dependence of the leakage current in hafnium oxide can be explained in terms of a microscopic model for TAT with temperature dependent capture cross section, where the relevant traps are uniformly distributed in the layer volume at 1.6 eV below the hafnium oxide conduction band. The cross section has an Arrhenius-like behavior with a rather high activation energy of 0.542 eV. For the samples of $HfSi_XO_YN_Z$ at hand, we observe a much smaller dependence of the leakage current on temperature, that can be simply explained by considering only the pure tunneling component.

## VIII. Acknowledgements

This work was supported in part by the SINANO Network of Excellence and by the MIUR-PRIN project: "Alternative Models and Architectures for nanoMOSFETs" . The authors wish to thank G. Groeseneken and IMEC for providing the samples used in this work.

Figure 1: Band diagram and 2D subband energies obtained from the 1D Poisson-Schrödinger solver for structure *a* (4 nm of $HfO_2$ and 1 nm of $SiO_2$) with an applied gate voltage of 3V.

Figure 2: Experimental and Theoretical Capacitance-Voltage characteristics (with and without Fermi level pinning) for structure *a* (4 nm of $HfO_2$ and 1 nm of $SiO_2$).

Figure 3: Experimental and theoretical Capacitance-Voltage characteristics (with and without Fermi level pinning) for structure *b* (2 nm of $HfSi_XO_YN_Z$ and 1 nm of SiON).

Figure 4: Experimental and theoretical Current-Voltage characteristics for structures *a* (4 nm of $HfO_2$ and 1 nm of $SiO_2$) and *c* (1 nm of $HfSi_XO_YN_Z$ and 1 nm of SiON).

Figure 5: J-V comparison for different MOSFET channel length. At low negative voltage the gate current density increases in shorter channel length MOSFET.

Figure 6: Current density as a function of the gate voltage at different temperatures (from 298 to 473 K with a step of 25 K) for structure *a* (4 nm of $HfO_2$ and 1 nm of $SiO_2$). Reproduced from figure 7 of Ref. [17]

Figure 7: Current density as a function of the gate voltage at different temperatures (from 298 to 448 K with a step of 25 K) for structure *b* (2 nm of $HfSi_XO_YN_Z$ and 1 nm of SiON).

Figure 8: Band profile of structure *a* (4 nm of $HfO_2$ and 1 nm of $SiO_2$) with a positive gate applied voltage of 0.5 V. In figure is also indicated the generation and recombination rates and a trap state $|\alpha\rangle$ in the hafnium oxide bulk.

Figure 9: Band profile of structure *a* (4 nm of $HfO_2$ and 1 nm of $SiO_2$) with a positive gate applied voltage of $\underline{1}$ V as obtained by a Poisson-Schrödinger simulation. The thicker line indicates the trap level 1.6 eV below the hafnium oxide conduction band.

Figure 10: current density as a function of the gate voltage for various half width of the Lorentzian function. The best fit is reached for 0.081 eV of the half width.

Figure 11: Extracted capture cross section as a function of temperature and Arrhenius fitting curve. In figure are also reported the activation energy and the capture cross section for $T \to \infty$

Figure 12: Theoretical and experimental curves for positive gate applied voltage at various temperatures for structure *a* (4 nm of $HfO_2$ and 1 nm of $SiO_2$).

Figure 13: Band profile of structure *a* (4 nm of $HfO_2$ and 1 nm of $SiO_2$) with a negative gate applied voltage of -1.5 V. With a thicker line is represented the trap band 1.6 eV below the hafnium oxide conduction band.

Figure 14: Theoretical and experimental curves for negative gate applied voltage at various temperatures for structure *a* (4 nm of $HfO_2$ and 1 nm of $SiO_2$).

Figure 15: Theoretical and experimental curves for positive gate applied voltage at various temperatures for structure *c* (1 nm of $HfSi_XO_YN_Z$ and 1 nm of SiON) at 298 and 400 K.

TABLE I
GATE STACK STRUCTURES CONSIDERED

|  | Structure *a* | Structure *b* | Structure *c* |
|---|---|---|---|
| high-*k* material | $HfO_2$ | $HfSi_XO_YN_Z$ | $HfSi_XO_YN_Z$ |
| $t_{high\text{-}k}$ | 4 nm | 2 nm | 1 nm |
| $t_{int}$ | 1nm ($SiO_2$) | 1 nm (SiON) | 1 nm (SiON) |
| EOT | 1.63 nm | 1.7 nm | 1.35 nm |

**Table 1**
**Andrea Campera, Giuseppe Iannaccone and Felice Crupi**

**IEEE Trans. Electron Devices**

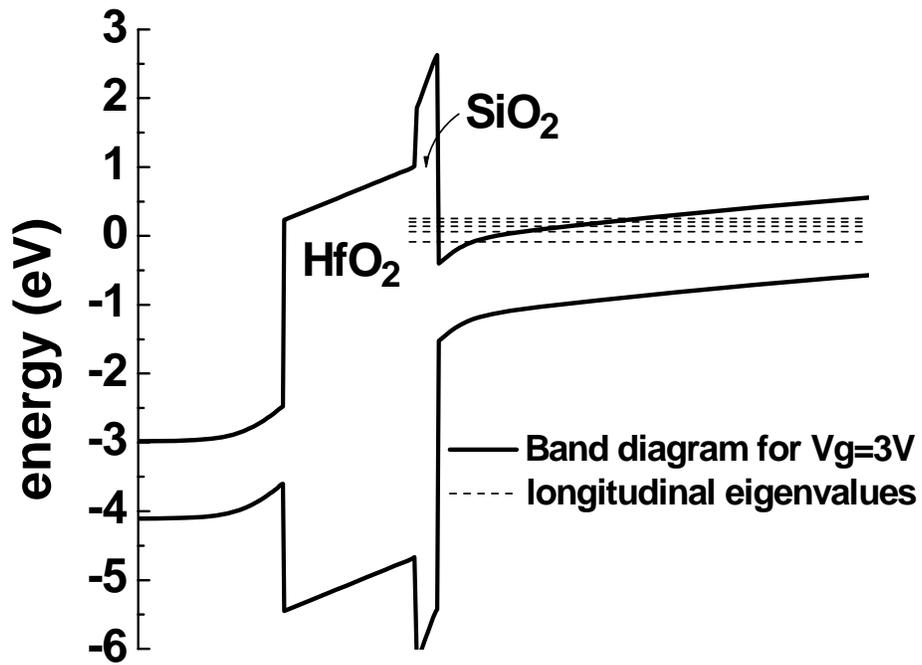

**Figure 1**
Andrea Campera, Giuseppe Iannaccone and Felice Crupi

**IEEE Trans. Electron Devices**

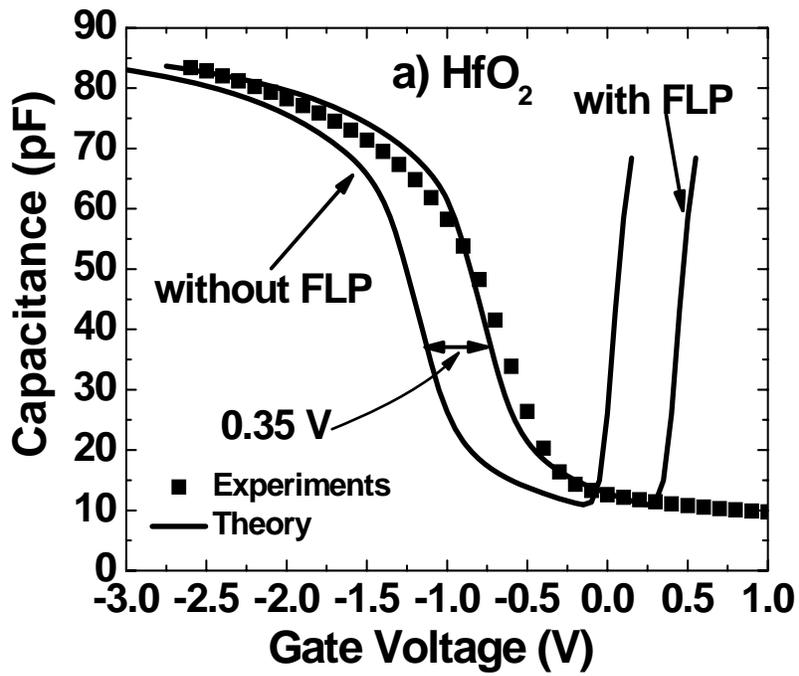

**Figure 2**
**Andrea Campera, Giuseppe Iannaccone and Felice Crupi**

**IEEE Trans. Electron Devices**

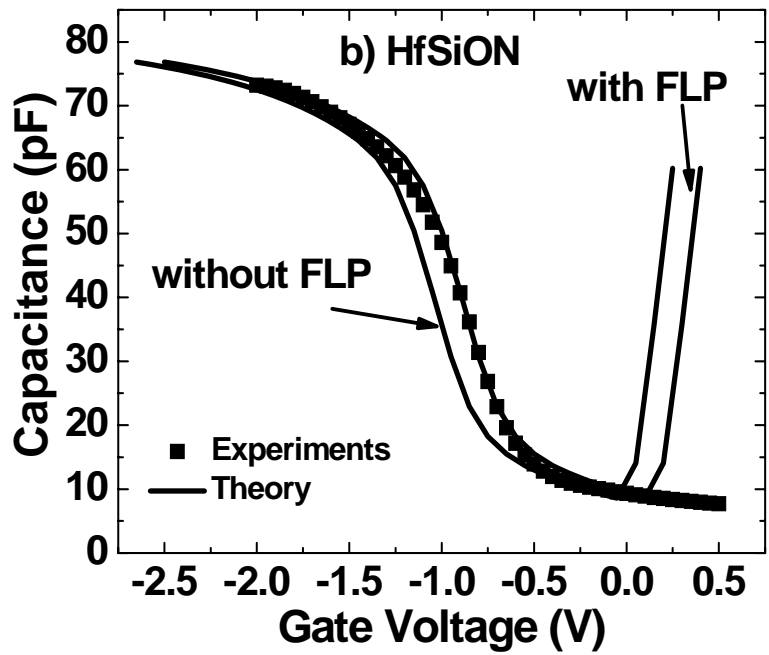

**Figure 3**
Andrea Campera, Giuseppe Iannaccone and Felice Crupi

**IEEE Trans. Electron Devices**

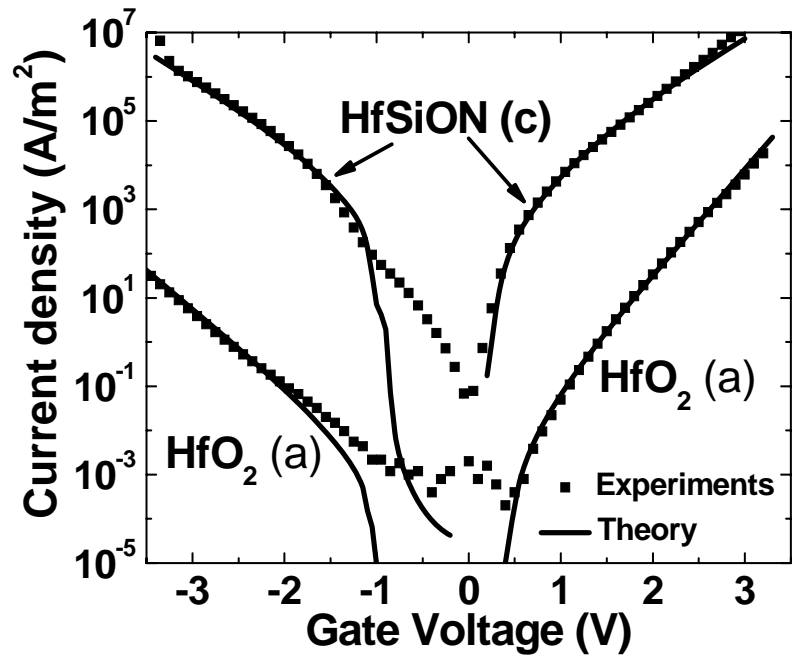

**Figure 4**
Andrea Campera, Giuseppe Iannaccone and Felice Crupi

**IEEE Trans. Electron Devices**

TABLE II
SUMMARY OF PHYSICAL PARAMETERS EXTRACTED FOR $HfO_2$, $HfSiON$ AND $SiON$

|  | $HfO_2$ | HfSiON | SiON |
|---|---|---|---|
| *electron affinity* | 1.75 eV | 1.97 eV | 1.27 eV |
| *electron effective mass* | 0.08 $m_0$ | 0.24 $m_0$ | 0.45 $m_0$ |
| *relative dielectric constant* | 25 | 11 | 5 |
| *FLP* | 0.35 V | 0.13 V | - |

**Table 2**
**Andrea Campera, Giuseppe Iannaccone and Felice Crupi**

**IEEE Trans. Electron Devices**

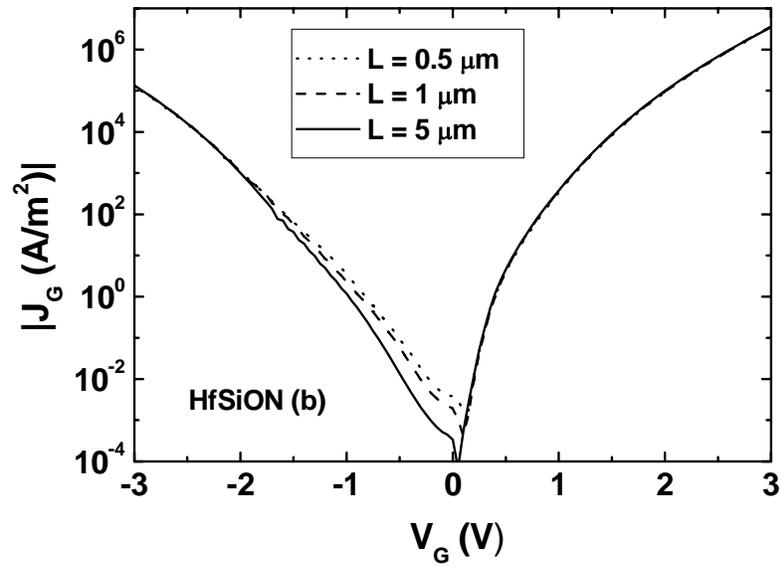

**Figure 5**
Andrea Campera, Giuseppe Iannaccone and Felice Crupi

**IEEE Trans. Electron Devices**

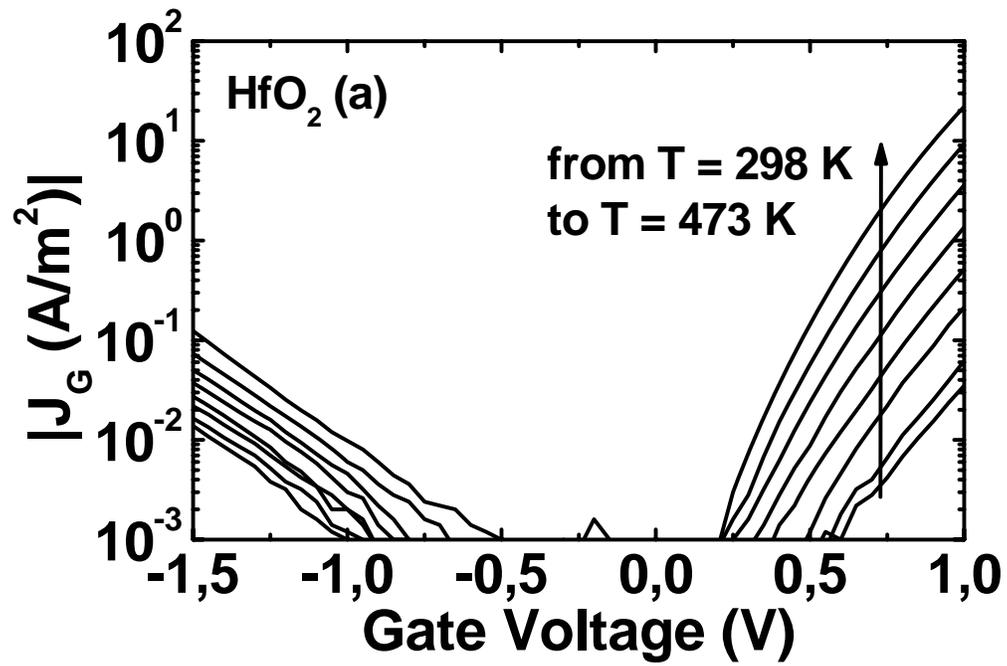

**Figure 6**
Andrea Campera, Giuseppe Iannaccone and Felice Crupi

**IEEE Trans. Electron Devices**

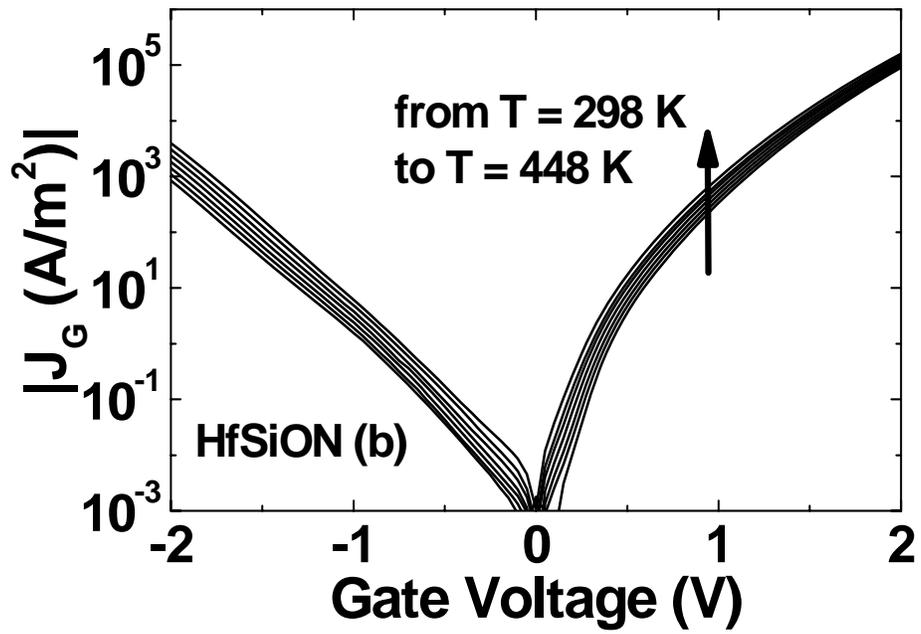

**Figure 7**
Andrea Campera, Giuseppe Iannaccone and Felice Crupi

IEEE Trans. Electron Devices

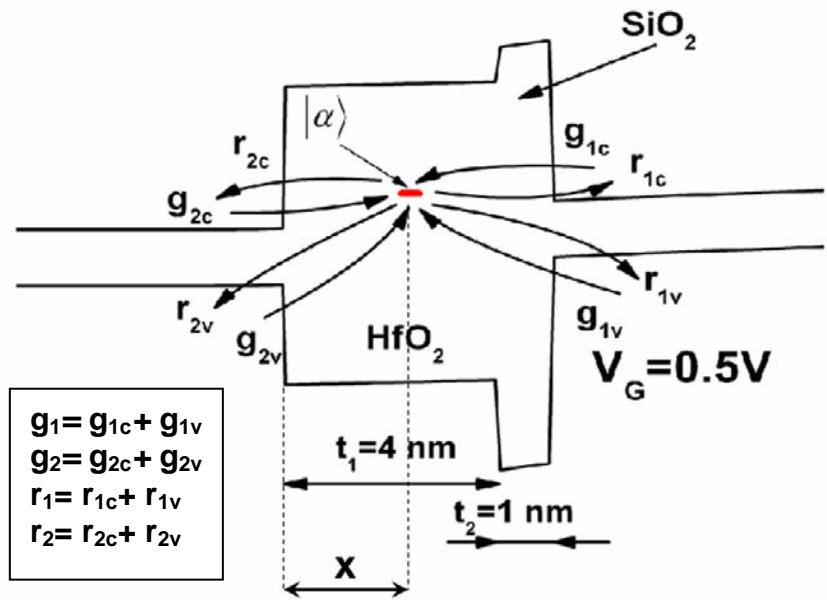

**Figure 8**
Andrea Campera, Giuseppe Iannaccone and Felice Crupi

**IEEE Trans. Electron Devices**

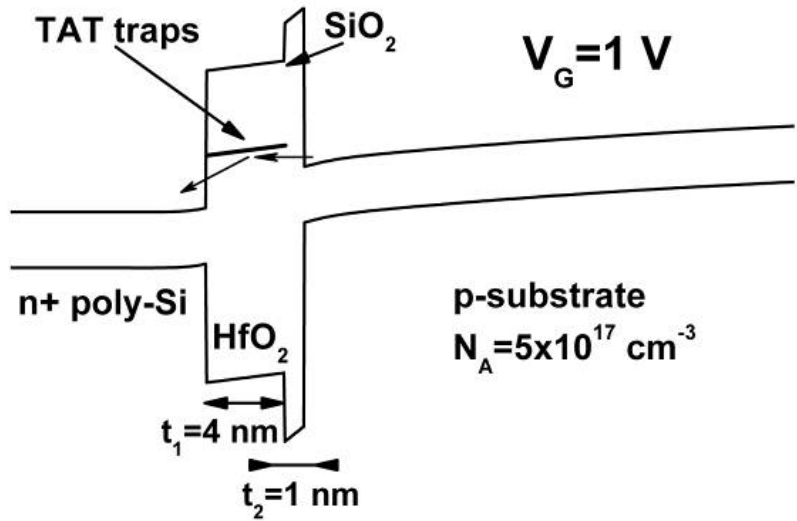

**Figure 9**
Andrea Campera, Giuseppe Iannaccone and Felice Crupi

**IEEE Trans. Electron Devices**

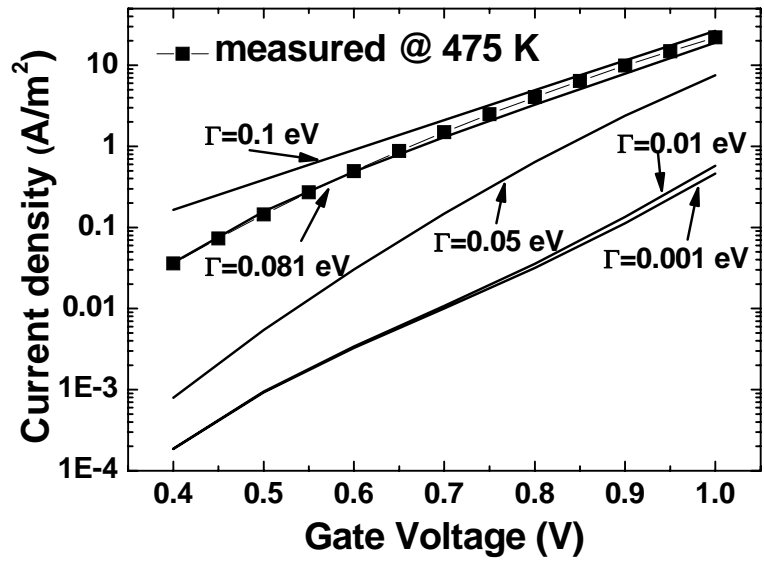

**Figure 10**
**Andrea Campera, Giuseppe Iannaccone and Felice Crupi**

**IEEE Trans. Electron Devices**

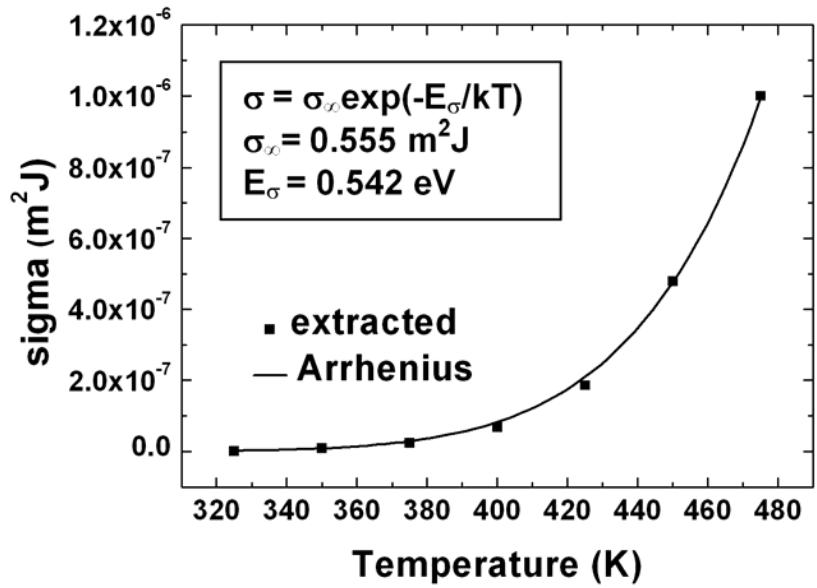

**Figure 11**
**Andrea Campera, Giuseppe Iannaccone and Felice Crupi**

**IEEE Trans. Electron Devices**

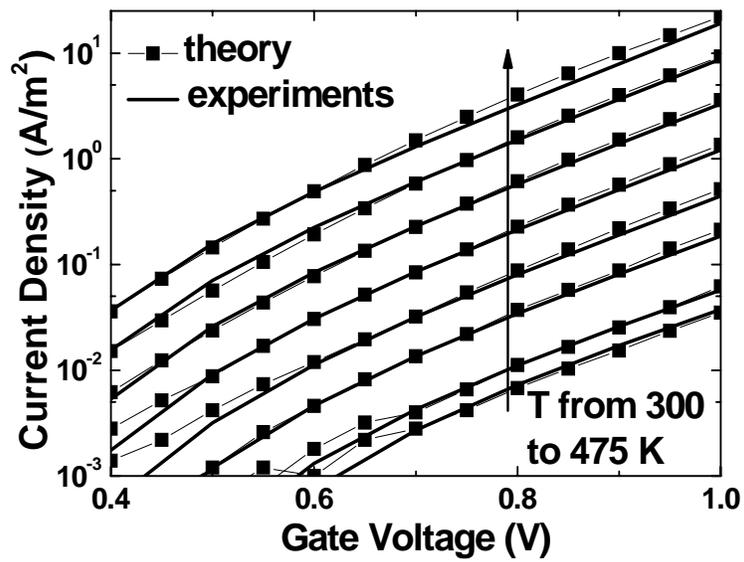

**Figure 12**
**Andrea Campera, Giuseppe Iannaccone and Felice Crupi**

**IEEE Trans. Electron Devices**

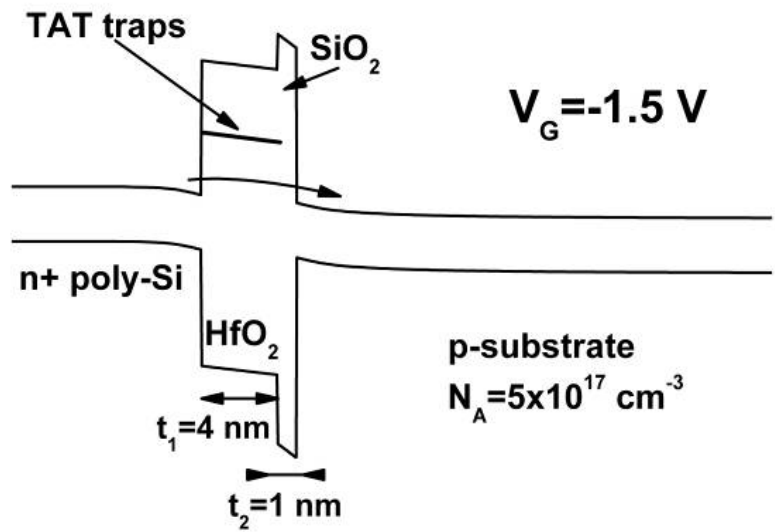

**Figure 13**
Andrea Campera, Giuseppe Iannaccone and Felice Crupi

**IEEE Trans. Electron Devices**

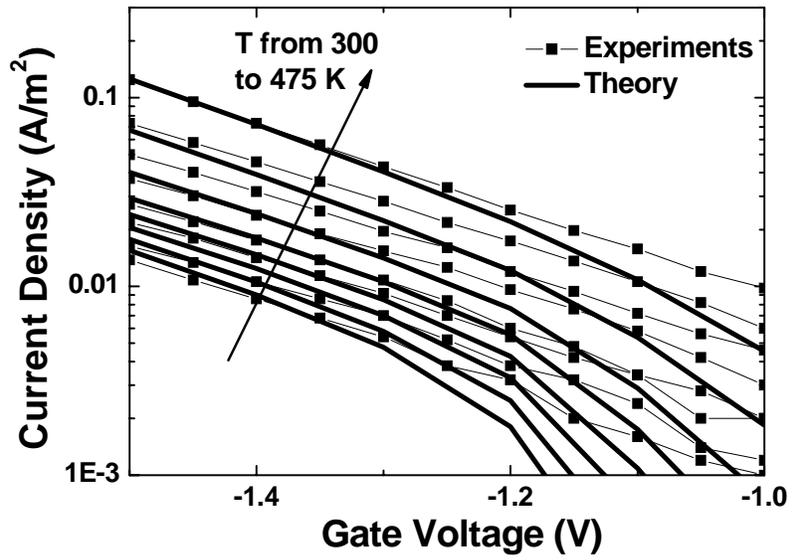

**Figure 14**
**Andrea Campera, Giuseppe Iannaccone and Felice Crupi**

**IEEE Trans. Electron Devices**

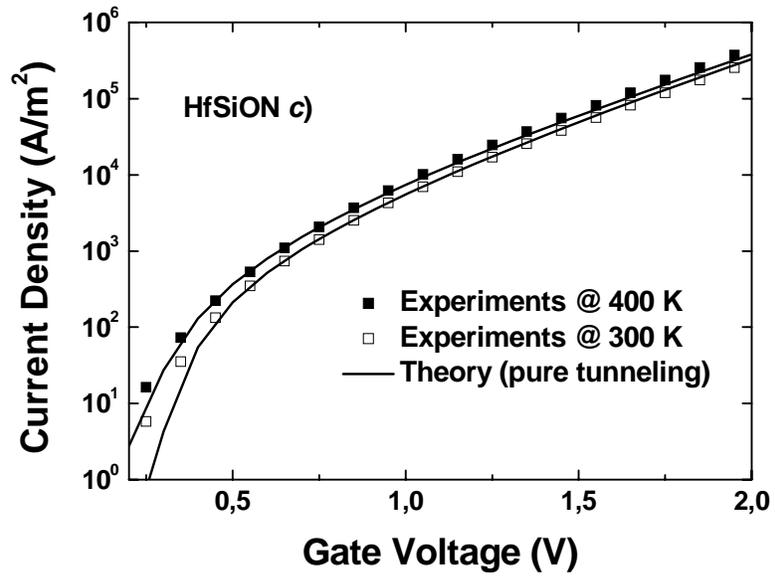

**Figure 15**
**Andrea Campera, Giuseppe Iannaccone and Felice Crupi**

**IEEE Trans. Electron Devices**